# Multi-Level Visual Similarity Based Personalized Tourist Attraction Recommendation Using Geo-Tagged Photos

Ling Chen*

Zhejiang University

Dandan Lyu

Zhejiang University

Shanshan Yu

Zhejiang University

Gencai Chen

Zhejiang University

Geo-tagged photo based tourist attraction recommendation can discover users' travel preferences from their taken photos, so as to recommend suitable tourist attractions to them. However, existing visual content based methods cannot fully exploit the user and tourist attraction information of photos to extract visual features, and do not differentiate the significance of different photos. In this paper, we propose multi-level visual similarity based personalized tourist attraction recommendation using geo-tagged photos (MEAL). MEAL utilizes the visual contents of photos and interaction behavior data to obtain the final embeddings of users and tourist attractions, which are then used to predict the visit probabilities. Specifically, by crossing the user and tourist attraction information of photos, we define four visual similarity levels and introduce a corresponding quintuplet loss to embed the visual contents of photos. In addition, to capture the significance of different photos, we exploit the self-attention mechanism to obtain the visual representations of users and tourist attractions. We conducted experiments on two datasets crawled from Flickr, and the experimental results proved the advantage of this method.

## 1 INTRODUCTION

With the advent of the smart era, people can easily share their travel experiences on social platforms, e.g., uploading some wonderful photos during a trip, forming abundant geo-tagged photos [1-3]. Users can manually search through the miscellaneous online information to find a few tourist attractions that meet their travel preferences. This usually costs much time and energy. Tourist attraction recommendation systems [4] can provide users with great convenience, as it can infer their travel preferences from travel history and automatically plan their trips.

* This work was funded by the National Key Research and Development Program of China (No. 2018YFB0505000) and the Fundamental Research Funds for the Central Universities (No. 2020QNA5017).
Authors' addresses: L. Chen (Corresponding author), College of Computer Science and Technology, Alibaba-Zhejiang University Joint Research Institute of Frontier Technologies, Zhejiang University, Hangzhou 310027, China; email: lingchen@cs.zju.edu.cn; D. Lyu, College of Computer Science and Technology, Zhejiang University, Hangzhou 310027, China; email: revaludo@cs.zju.edu.cn; S. Yu, College of Computer Science and Technology, Zhejiang University, Hangzhou 310027, China; email: shshyu@cs.zju.edu.cn; G. Chen, College of Computer Science and Technology, Zhejiang University, Hangzhou 310027, China; email: chengc@cs.zju.edu.cn.

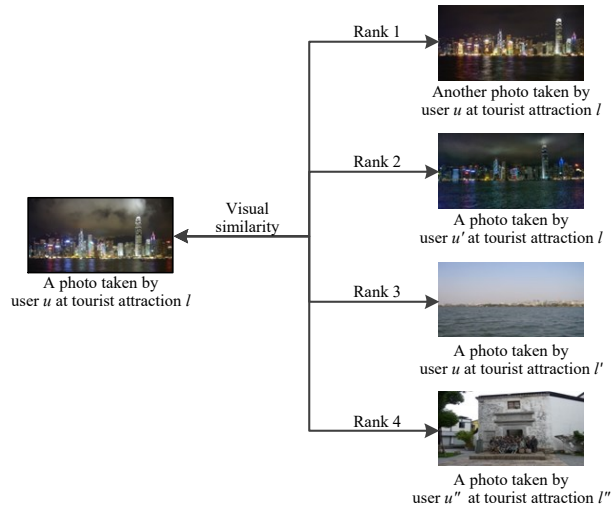

Figure 1: An illustration of multi-level visual similarity.

In the past decade, geo-tagged photo based tourist attraction recommendation has become one of the research hotspots. In the early years, researchers mainly consider users' travel preferences to make personalized recommendation [5, 6]. In recent years, various types of side information have been introduced to get more appropriate recommendation results [7-9]. Owing to the efficiency and effectiveness of deep neural networks (DNNs) in image processing, the visual contents of photos have gradually received attention. Existing visual content based methods usually first extract features from the visual contents of photos, and then use these features as prior knowledge to constrain the training of the recommendation model constructed based on users' travel history [10-12]. These methods have some drawbacks: 1) They treat different photos taken by a user or taken at a tourist attraction equally via average or max pooling the visual features of photos [11, 12], without differentiating their significance. 2) They cannot extract visual features adaptive to tourist attraction recommendation, as the extraction of visual features is mostly guided by computer vision tasks that have no relationship with the recommendation scenario.

A method named VPOI [13] that can extract visual features adaptive to tourist attraction recommendation has been proposed. It jointly extracts features from the visual contents of photos, classifies the photos according to who they are taken by and where they are taken, and factorizes the user-tourist attraction interaction matrix for personalized recommendation. Given a photo, this method independently exploits the user and tourist attraction information to partition other photos into visually similar/non-similar groups, assuming that the similarities of the photos taken by the same user or taken at the same tourist attraction are higher than those of the other photos. However, this method cannot capture multi-level visual similarity as shown in Figure 1, i.e., given a photo, its visual similarity with another photo taken by the same user at the same tourist attraction ranks first, as they usually capture same objects from different directions; its visual similarity with another photo taken by a different user at the same tourist attraction ranks second, as they usually capture same objects from different directions and under different lighting conditions; its visual similarity with another photo taken by the same user at a different tourist attraction ranks third, as they usually capture similar sceneries that the user prefers; its visual similarity with another photo taken by a different user at a different tourist attraction ranks last, as the travel preferences of different users and the sceneries of different tourist attractions usually differ greatly.

To deal with the aforementioned problems, we propose multi-level visual similarity based personalized tourist attraction recommendation using geo-tagged photos (MEAL). By crossing the user and tourist attraction information of photos, we define multi-level similarity for visual content embedding.

The crucial contributions of this paper are summarized as below:

1) Propose MEAL, combining the visual representations obtained by fusing the visual features of photos through the self-attention mechanism and the latent factors obtained by factorizing the user-tourist attraction interaction matrix to obtain the final embeddings of users and tourist attractions, which can capture the significance of different photos for representing users and tourist attractions.

2) Propose multi-level similarity aware visual content embedding for geo-tagged photos, trying to ensure that, considering the visual similarities with a given photo, other photos are ranked as follows: photos taken by the same user at the same tourist attraction > photos taken by different users at the same tourist attraction > photos taken by the same user at different tourist attractions > photos taken by different users at different tourist attractions, which can fully exploit the user and tourist attraction information of photos.

3) Evaluate the proposed method on two real-world datasets crawled from Flickr and make comparison with the state-of-the-art methods. The experimental results show the advantage of this method.

The rest of this paper is organized as follows. Section 2 reviews the related work. Section 3 gives the preliminaries of this paper and defines the research problem. Section 4 introduces the proposed method MRATE in detail. Section 5 presents the experimental settings and results. Finally, Section 6 concludes the paper and gives a brief discussion of the future work.

## 2 RELATED WORK

In this part, some recent works closely related to our work are introduced, consisting of geo-tagged photo based tourist attraction recommendation and deep metric learning based visual content embedding.

### 2.1 Geo-Tagged Photo Based Tourist Attraction Recommendation

Geo-tagged photos imply the travel history of users, which provide rich data for tourist attraction recommendation. In the early years, researchers mainly consider users' travel preferences to make personalized recommendation [5, 6]. Clements et al. [5] firstly computed the similarities between users based on the Gaussian kernel convolution values of their geotag distributions in a common visited city, and then recommended tourist attractions in a previously unvisited city according to the rankings of users with similar travel preferences. Popescu and Grefenstette [6] also followed the idea of collaborative filtering, but they applied different similarity measures compared to [5].

To get more appropriate recommendation results, various types of side information have been introduced [7-9]. Majid et al. [8, 9] proposed to recommend tourist attractions and tourist routes by considering users' travel preferences under different contexts (e.g., season and weather). Bhargava et al. [7] jointly factorized user-tourist attraction-activity-time tensor, tourist attraction-activity matrix, tourist attraction-tourist attraction similarity matrix, and activity-activity correlation matrix to provide multi-dimensional recommendation.

Owing to the efficiency and effectiveness of DNNs in image processing, the visual contents of photos have gradually received attention. Existing visual content based methods usually first extract features from the visual contents of photos, and then use these features as prior knowledge to constrain the training of the recommendation model constructed based on users' travel history [10-12]. DTMMF [10] firstly extracted the gender and age information of people appearing in photos to represent users and tourist attractions, based on which user-user and tourist attraction-tourist attraction similarities were calculated to constrain the factorization of the user-tourist attraction interaction matrix. WIND-MF [11] followed a similar idea; one of the main differences is that it extracted the visual feature of each photo via a variational auto-encoder, and then averaged the visual features of corresponding photos to get the visual representations of users and tourist attractions. VPMF [12] extracted more visual features compared to WIND-MF [11], including the color histogram features, scale-invariant feature transform (SIFT) features, and VGG16 features extracted via a pre-trained network, and then obtained the visual representations of users and tourist attractions via max pooling. These methods treat different photos taken by a user or taken at a tourist attraction equally via average or max pooling the visual features of photos [11, 12], without differentiating their significance. In order to capture the significance of different photos for representing users and tourist attractions, we introduce the self-attention mechanism to infer the weights of photos.

In addition, existing visual content based methods cannot extract visual features adaptive to tourist attraction recommendation, as the extraction of visual features is mostly guided by computer vision tasks that have no relationship with the recommendation scenario. To deal with this problem, VPOI [13] was proposed, which jointly extracts features from the visual contents of photos via a VGG16 model, classifies the photos according to who they are taken by and where they are taken, and factorizes the user-tourist attraction interaction matrix for personalized recommendation. Specifically, the visual feature of a photo and the factorized latent vector of a user are fed into a softmax function to identify the probability that the photo is taken by the user. Similarly, the visual feature of a photo and the factorized latent vector of a tourist attraction are fed into a softmax function to identify the probability that the photo is taken at the tourist attraction. Given a photo, this method independently exploits the user and tourist attraction information to partition other photos into visually similar/non-similar groups, assuming that the similarities of the photos taken by the same user or taken at the same tourist attraction are higher than those of the other photos. However, the photos taken by the same user at different tourist attractions may vary significantly on visual features, as the sceneries of different tourist attractions are usually different, and the photos taken at the same tourist attraction by different users may also vary significantly on visual features, as the preferences of different users are usually different, which cannot be captured by this method. In order to better capture the preferences of users and the sceneries of tourist attractions, we introduce multi-level similarity to extract visual features adaptive to tourist attraction recommendation.

## 2.2 Deep Metric Learning Based Visual Content Embedding

Triplet loss based deep metric learning methods have been widely applied in computer vision and pattern recognition area [14-16]. Zeng et al. [14] proposed Hierarchical Clustering with hard-batch Triplet loss (HCT), which makes full use of the similarity among samples in the target dataset through hierarchical clustering, reduces the influence of hard examples through hard-batch triplet loss, so as to generate high quality pseudo labels and improve model performance. Liao and Shao [15] proposed graph sampling (GS), which builds a nearest neighbor relationship graph for all classes at the beginning of each epoch. For each mini batch, GS randomly selects a class and its nearest neighboring classes so as to provide informative and challenging examples for learning. Zhou and Patel [16] proposed Hardness Manipulation to efficiently perturb the training triplet till a specified level of hardness for adversarial training, according to a harder benign triplet or a pseudo-hardness function. For these methods, triplet sampling is crucial for fast and stable convergence. In this paper, we employ semi-hard sampling [17], which converges more quickly while being less aggressive.

Modelling similarity at different levels can yield better classification results [19-21]. Yang et al. [19] proposed sentiment constraints for understanding affective images via deep metric learning, which considers emotion labels with the same or different polarities by generalizing the triplet loss. Zhang et al. [20] embedded label structures (e.g., hierarchy or shared attributes) by generalizing the triplet loss to obtain fine-grained feature representations. Inspired by these studies where multi-level similarity is defined by a tree-like hierarchy [20] or cluster distribution [21], we define four similarity levels by crossing the user and tourist attraction information of photos. A corresponding quintuplet loss is then introduced to ensure the proper order of these similarities when embedding the visual contents of photos.

## 3 PRELIMINARIES

In this part, we firstly formally define some basic concepts used throughout the paper, and then clarify the research problem of this paper.

**Notation:** Capital letters denote sets, and $|\ |$ denotes the cardinality of a set. Bold upper-case letters denote matrices, and bold lower-case letters denote vectors. $^T$ denotes the transpose operation. $\|\ \|_2^2$ denotes the Euclidean norm of a vector. $||$ denotes the vector concatenation operator.

**Definition 1:** (Geo-tagged photo) A geo-tagged photo is usually taken by a user at an interesting place during a trip, and contains time and geographical coordinate (usually referred as geotag) information indicating when and where it was taken. The geo-tagged photo set can be denoted by $P = \{p_1, p_2, \cdots, p_{|P|}\}$. The users taking these geo-tagged photos can be denoted by $U = \{u_1, u_2, \cdots, u_{|U|}\}$.

**Definition 2:** (Tourist attraction) A tourist attraction is a specific geographic area in a city, e.g., a park, a museum, and a lake, which is usually visited and photographed frequently by tourists, and can be denoted by $l = (c, g)$, where $c$ is the city it lies in and $g$ is its geographical coordinate. The tourist attraction set can be denoted by $L = \{l_1, l_2, \cdots, l_{|L|}\}$. The cities containing these tourist attractions can be denoted by $C = \{c_1, c_2, \cdots, c_{|C|}\}$.

**Definition 3:** (Visit) A visit indicating that at time $t$, tourist attraction $l$ is visited by user $u$ can be denoted by $v = (u, l, t)$.

**Definition 4:** (User-tourist attraction interaction matrix) A user-tourist attraction interaction matrix indicating the visit frequencies of users to tourist attractions can be denoted by $\mathbf{M} \in \mathrm{R}^{|U| \times |L|}$.

**Definition 5:** (Quintuplet) A quintuplet can be denoted by $e = (p_o, p_o^{++}, p_o^{-+}, p_o^{+-}, p_o^{--})$, where $p_o \in P$ is a geo-tagged photo taken by a specific user at a specific tourist attraction, $p_o^{++} \in P$ denotes another photo taken by the same user and at the same tourist attraction as $p_o$, $p_o^{-+} \in P$ denotes a photo taken by a different user from $p_o$ but at the same tourist attraction as $p_o$, $p_o^{+-} \in P$ denotes a photo taken by the same user as $p_o$ but at a different tourist attraction from $p_o$, $p_o^{--} \in P$ denotes a photo taken by a different user and at a different tourist attraction from $p_o$. The quintuplet set can be denoted by $Q = \{e_1, e_2, \cdots, e_{|Q|}\}$.

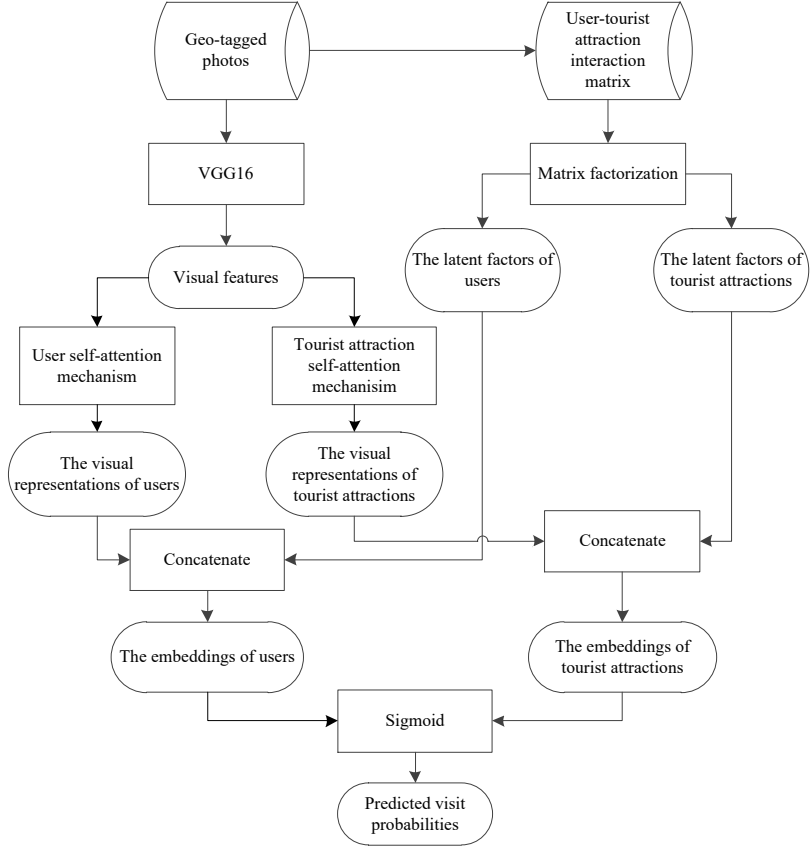

Figure 2: The framework of MEAL.

The research problem of this paper is: Given the geo-tagged photos $P$ taken by users $U$ in cities $C$, for a user $u \in U$ and a city $c \in C$ where the user has never visited, i.e., the query is $q = (u, c)$, we want to recommend a list of tourist attractions in city $c$ that user $u$ would be interested in.

## 4 METHODOLOGY

Figure 2 shows the framework of MEAL. Firstly, we extract user-tourist attraction interaction matrix from geo-tagged photos. Then, we extract the visual features of photos via the VGG16 model, based on which we utilize the self-attention mechanism to obtain the visual representations of users and tourist attractions. We also factorize the user-tourist attraction interaction matrix to obtain the latent factors of users and tourist attractions. Afterwards, we concatenate the visual representations and latent factors to obtain the final embeddings of users and tourist attractions, based on which we can predict the visit probabilities.

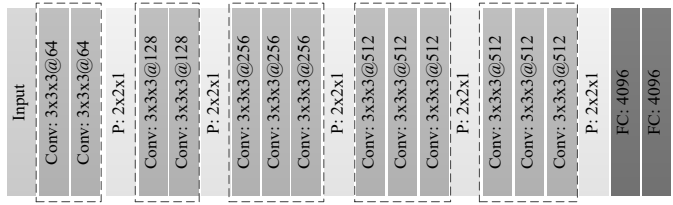

Figure 3: The architecture of the VGG16 model for visual feature extraction. "Conv", "P", and "FC" denote convolutional, pooling, and fully connected layers, respectively. The numbers before and after "@" denote the sizes and the numbers of convolutional kernels, respectively. The numbers of pooling layers denote the sizes of pooling ranges. The numbers of fully connected layers denote the numbers of neurons.

### 4.1 Extracting Interaction Matrix

Classical clustering algorithms, e.g., mean-shift and DBSCAN, have been exploited to extract tourist attractions from geo-tagged photos [22, 23]. P-DBSCAN [24] is a density-based clustering algorithm specialized for place analysis using large collections of geo-tagged photos, which defines neighborhood density as the number of users who have taken photos in the area, and proposes adaptive density to optimize search for dense areas. Specifically, by inputting the geographical coordinate and user information of geo-tagged photos to the P-DBSCAN algorithm, we can obtain the tourist attraction set $L$.

The preference score of a user to a tourist attraction is proportional to the corresponding visit frequency. Like Xu et al. [25], the geo-tagged photos taken by user $u_i$ at tourist attraction $l_j$ are firstly ordered according to their taken time. Secondly, visits are identified by considering the taken time difference between successive photos. Specifically, several successive photos are assumed to be taken within a same visit if the taken time difference between the beginning photo and the ending photo is smaller than visit duration threshold $t_{\text{thr}}$, as a user may have taken multiple geo-tagged photos within one visit. Then the time of this visit is calculated by averaging the taken time of these photos. Thirdly, we count the number of visits to obtain the visit frequency of user $u_i$ to tourist attraction $l_j$, i.e., $\mathbf{M}_{ij}$. After processing the geo-tagged photos of all the possible user-tourist attraction pairs, we can obtain the user-tourist attraction interaction matrix $\mathbf{M}$.

### 4.2 Extracting Visual Features

VGG16 [26] is a deep learning model designed for image classification task and has shown its efficiency in various tasks, e.g., video captioning [27, 28], multimedia retrieval [29, 30], and recommendation [12, 13], which can obtain representative visual features of input images. Specifically, given a geo-tagged photo $p_k$, its pixel values are firstly resized into a tensor of shape 224×224×3, which is then input to a VGG16 model to extract its visual feature, denoted by $\mathbf{v}_k \in \mathrm{R}^d$.

The architecture of the VGG16 model for visual feature extraction is illustrated in Figure 3, which is composed of five convolution blocks and five corresponding pooling layers, as well as two fully connected layers. The first two convolution blocks contain two convolutional layers, while the last three ones contain three convolutional layers. Specifically, the output dimension of the VGG16 model is decided by the number of neurons of the last fully connected layer, i.e., $d = 4096$.

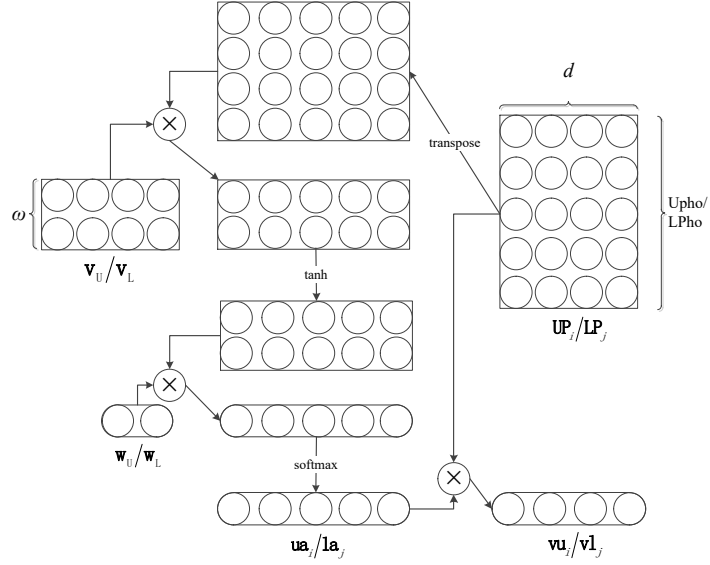

Figure 4: The self-attention mechanism based visual representation learning.

### 4.3 Obtaining Visual Representations

The attention mechanism has shown its efficiency in various tasks, e.g., image classification [31, 32], image captioning [33, 34], and recommendation [35, 36], and it can urge a model to concentrate on a specific part of features at a time. In order to capture the significance of different photos for a user or a tourist attraction, the attention mechanism is introduce here to infer the weights of photos. Specifically, by the self-attention mechanism [37], the weights of the photos taken by a same user or taken at a same tourist attraction is firstly calculated based on their visual features, and then the respective visual representation is obtained by computing the weighted sum of corresponding visual features, which is shown in Figure 4.

The weights assigned by the self-attention mechanism can be calculated by (1) and (2).

$$\mathbf{ua}_i = \mathrm{softmax}(\mathbf{w}_U \tanh(\mathbf{V}_U \mathbf{UP}_i^\mathrm{T})) \tag{1}$$

$$\mathbf{la}_j = \mathrm{softmax}(\mathbf{w}_L \tanh(\mathbf{V}_L \mathbf{LP}_j^\mathrm{T})) \tag{2}$$

where $\mathbf{UP}_i \in \mathrm{R}^{\mathrm{UPho} \times d}$ and $\mathbf{LP}_j \in \mathrm{R}^{\mathrm{LPho} \times d}$ are formed by vertically stacking the visual features of photos taken by user $u_i$ and taken at tourist attraction $l_j$ in an ascending order of their taken time, and UPho and LPho are the max numbers of photos taken by a user and taken at a tourist attraction, respectively. Note that, if user $u_i$ or tourist attraction $l_j$ has fewer than UPho or LPho photos, we vertically stack the visual features of these photos and multiple mean vectors of these visual features to form $\mathbf{UP}_i$ or $\mathbf{LP}_j$. $\mathbf{w}_U$ and $\mathbf{w}_L$ are learnable weight vectors with length $\omega$. $\mathbf{V}_U \in \mathrm{R}^{\omega \times d}$ and $\mathbf{V}_L \in \mathrm{R}^{\omega \times d}$ are learnable weight matrices. $\mathbf{ua}_i$ is the attentive weight vector of $\mathbf{UP}_i$ with length UPho. $\mathbf{la}_j$ is the attentive weight vector of $\mathbf{LP}_j$ with length LPho. The softmax function ensures the entire computed weights sum up to 1.

Then, we sum up the vectors in $\mathbf{UP}_i$ and $\mathbf{LP}_j$ according to the weights provided by $\mathbf{ua}_i$ and $\mathbf{la}_j$ to get the visual representation of user $u_i$, i.e., $\mathbf{vu}_i$, and the visual representation of tourist attraction $l_j$, i.e., $\mathbf{vl}_j$, by (3) and (4), respectively.

$$\mathbf{vu}_i = \mathbf{ua}_i \mathbf{UP}_i \tag{3}$$

$$\mathbf{vl}_j = \mathbf{la}_j \mathbf{LP}_j \tag{4}$$

### 4.4 Obtaining Latent Factors

Matrix factorization has been widely applied in recommender systems to model user-item interactions [38-41], which can project users and items into a same low dimensional latent space. Weighted matrix factorization [42] is exploited here to model visit frequency, which is a kind of implicit feedback, and the objective function is given by (5).

$$V = \min \frac{1}{2} \sum_{0<i \leq |U|, 0<j \leq |L|} \mathbf{C}_{ij} \left(\mathbf{R}_{ij} - \mathbf{hu}_i \mathbf{hl}_j^{\mathrm{T}}\right)^2 + \frac{\lambda_1}{2} \left( \sum_{0<i \leq |U|} \|\mathbf{hu}_i\|_2^2 + \sum_{0<j \leq |L|} \|\mathbf{hl}_j\|_2^2 \right) \tag{5}$$

where $\mathbf{hu}_i$ and $\mathbf{hl}_j$ are the latent factors of user $u_i$ and tourist attraction $l_j$, respectively. $\lambda_1$ is a hyper-parameter used to control the weights of regularization terms. $\mathbf{C}_{ij}$ denotes the confidence weight of $\mathbf{M}_{ij}$, and is formalized by (6). $\mathbf{R}_{ij} = 1$, if $\mathbf{M}_{ij} > 0$; otherwise $\mathbf{R}_{ij} = 0$.

$$\mathbf{C}_{ij} = 1 + \gamma \mathbf{M}_{ij} \tag{6}$$

where $\gamma$ is a hyper-parameter used to control the confidence weight of $\mathbf{M}_{ij}$.

### 4.5 Obtaining Final Embeddings

In order to combine both the visual contents of photos and interaction behavior data to represent users and tourist attractions, we obtain the final embedding of user $u_i$, i.e., $\mathbf{u}_i$, and the final embedding of tourist attraction $l_j$, i.e., $\mathbf{l}_j$, by concatenating the respective visual representation and latent factor, which can be formulated by (7) and (8), respectively.

$$\mathbf{u}_i = \mathbf{va}_i \| \mathbf{hu}_i \tag{7}$$

$$\mathbf{l}_j = \mathbf{vl}_j \| \mathbf{hl}_j \tag{8}$$

Note that, there are multiple alternative approaches can be utilized to fuse visual representation and latent factor into a single vector, we choose concatenation here, as it can reduce information loss and the number of model parameters.

### 4.6 Learning

#### 4.6.1 Predicting Visit Probabilities.

Finally, the probability that user $u_i$ will visit tourist attraction $l_j$ can be calculated by (9).

$$\widetilde{\mathbf{R}}_{ij} = \mathrm{sigmoid}(\mathbf{WP}(\mathbf{u}_i \| \mathbf{l}_j) + \mathbf{bp}) \tag{9}$$

where the sigmoid function ensures the prediction score to be in the range of [0, 1]. $\mathbf{WP}$ is a learnable weight matrix, and $\mathbf{bp}$ is a learnable bias vector.

The prediction loss is defined as the binary cross-entropy loss between the predicted visit probabilities and the ground truth, which can be formulated by (10).

$$L_{\mathrm{prediction}} = - \sum_{1 \leq i \leq |U|, 1 \leq j \leq |L|} \mathbf{R}_{ij} \log(\widetilde{\mathbf{R}}_{ij}) + (1 - \mathbf{R}_{ij}) \log(1 - \widetilde{\mathbf{R}}_{ij}) \tag{10}$$

#### 4.6.2 Preserving Multi-Level Visual Similarity.

Given a quintuplet $e = (p_o, p_o^{++}, p_o^{-+}, p_o^{+-}, p_o^{--})$, (11)-(19) need to be satisfied for preserving multi-level visual similarity.

$$\|\mathbf{v}_o - \mathbf{v}_o^{++}\|_2^2 + m_1 < \|\mathbf{v}_o - \mathbf{v}_o^{-+}\|_2^2 \tag{11}$$

$$\|\mathbf{v}_o - \mathbf{v}_o^{++}\|_2^2 + m_2 < \|\mathbf{v}_o - \mathbf{v}_o^{+-}\|_2^2 \tag{12}$$

$$\|\mathbf{v}_o - \mathbf{v}_o^{++}\|_2^2 + m_3 < \|\mathbf{v}_o - \mathbf{v}_o^{--}\|_2^2 \tag{13}$$

$$\|\mathbf{v}_o - \mathbf{v}_o^{-+}\|_2^2 + m_4 < \|\mathbf{v}_o - \mathbf{v}_o^{+-}\|_2^2 \tag{14}$$

$$\|\mathbf{v}_o - \mathbf{v}_o^{-+}\|_2^2 + m_5 < \|\mathbf{v}_o - \mathbf{v}_o^{--}\|_2^2 \tag{15}$$

$$\|\mathbf{v}_o - \mathbf{v}_o^{+-}\|_2^2 + m_6 < \|\mathbf{v}_o - \mathbf{v}_o^{--}\|_2^2 \tag{16}$$

$$0 < m_1 < m_2 < m_3 \tag{17}$$

$$0 < m_4 < m_5 \tag{18}$$

$$0 < m_6 \tag{19}$$

where $\mathbf{v}_o$, $\mathbf{v}_o^{++}$, $\mathbf{v}_o^{-+}$, $\mathbf{v}_o^{+-}$, and $\mathbf{v}_o^{--}$ are the visual features of $p_o$, $p_o^{++}$, $p_o^{-+}$, $p_o^{+-}$, and $p_o^{--}$. $m_1$, $m_2$, $m_3$, $m_4$, $m_5$, and $m_6$ are hyper-parameters used to control the margins between photo pairs $(p_o, p_o^{++})$ and $(p_o, p_o^{-+})$, $(p_o, p_o^{++})$ and $(p_o, p_o^{+-})$, $(p_o, p_o^{++})$ and $(p_o, p_o^{--})$, $(p_o, p_o^{-+})$ and $(p_o, p_o^{+-})$, $(p_o, p_o^{-+})$ and $(p_o, p_o^{--})$, as well as $(p_o, p_o^{+-})$ and $(p_o, p_o^{--})$, respectively.

The triplet loss of (11) is given by (20).

$$L_1^e = [\|\mathbf{v}_o - \mathbf{v}_o^{++}\|_2^2 - \|\mathbf{v}_o - \mathbf{v}_o^{-+}\|_2^2 + m_1]_+ \tag{20}$$

where the value of [ ]$_+$ is the same as the value in [ ] if it is positive, otherwise it is 0. We can obtain the corresponding triplet losses $L_2^e$, $L_3^e$, $L_4^e$, $L_5^e$, and $L_6^e$ of (12-16) similarly.

The final quintuplet loss can be calculated by (21).

$$L_{\text{QUIN}} = \sum_{e \in Q} L_1^e + L_2^e + L_3^e + L_4^e + L_5^e + L_6^e \tag{21}$$

where $Q$ includes the training quintuplets that are selected by using the hard mining technique proposed by Schroff et al. [17]. Specifically, for any photo $p_o \in P$, all the other photos taken by the same user and at the same tourist attraction as $p_o$ should be selected as $p_o^{++}$. After selecting $p_o^{++}$, all the photos taken by a different user from $p_o$ but at the same tourist attraction as $p_o$, photos taken by the same user as $p_o$ but at a different tourist attraction from $p_o$, as well as photos taken by a different user and at a different tourist attraction from $p_o$ that satisfy (22)-(27) should be selected as $p_o^{-+}$, $p_o^{+-}$, and $p_o^{--}$, respectively.

$$\|\mathbf{v}_o - \mathbf{v}_o^{++}\|_2^2 < \|\mathbf{v}_o - \mathbf{v}_o^{-+}\|_2^2 < \|\mathbf{v}_o - \mathbf{v}_o^{++}\|_2^2 + m_1 \tag{22}$$

$$\|\mathbf{v}_o - \mathbf{v}_o^{++}\|_2^2 < \|\mathbf{v}_o - \mathbf{v}_o^{+-}\|_2^2 < \|\mathbf{v}_o - \mathbf{v}_o^{++}\|_2^2 + m_2 \tag{23}$$

$$\|\mathbf{v}_o - \mathbf{v}_o^{++}\|_2^2 < \|\mathbf{v}_o - \mathbf{v}_o^{--}\|_2^2 < \|\mathbf{v}_o - \mathbf{v}_o^{++}\|_2^2 + m_3 \tag{24}$$

$$\|\mathbf{v}_o - \mathbf{v}_o^{-+}\|_2^2 < \|\mathbf{v}_o - \mathbf{v}_o^{+-}\|_2^2 < \|\mathbf{v}_o - \mathbf{v}_o^{-+}\|_2^2 + m_4 \tag{25}$$

$$\|\mathbf{v}_o - \mathbf{v}_o^{-+}\|_2^2 < \|\mathbf{v}_o - \mathbf{v}_o^{--}\|_2^2 < \|\mathbf{v}_o - \mathbf{v}_o^{-+}\|_2^2 + m_5 \tag{26}$$

$$\|\mathbf{v}_o - \mathbf{v}_o^{+-}\|_2^2 < \|\mathbf{v}_o - \mathbf{v}_o^{--}\|_2^2 < \|\mathbf{v}_o - \mathbf{v}_o^{+-}\|_2^2 + m_6 \tag{27}$$

### 4.6.3 Joint Learning.

The final objective function of MEAL is given by (28).

$$L = L_{\text{QUIN}} + L_{\text{prediction}} + \lambda_2 \|\Theta\|_F^2 \tag{28}$$

where $\Theta$ denotes the learnable parameters of the model. $\lambda_2$ is a hyper-parameter used to control the weight of parameter regularization term. $\|\ \|_F^2$ denotes the Frobenius norm of a matrix.

Table 1: The statistics of the datasets

| Dataset | City | # Photos | # Users | # Tourist attractions |
|---|---|---|---|---|
| Dataset1 | Beijing | 220,626 | 294 | 414 |
| | Chengdu | 18,513 | 68 | 49 |
| | Guangzhou | 16,999 | 40 | 55 |
| | Hangzhou | 28,192 | 133 | 94 |
| | Hong Kong | 185,003 | 210 | 419 |
| | Shanghai | 230,563 | 314 | 483 |
| Dataset2 | Barcelona | 5,203 | 174 | 50 |
| | Berlin | 5,614 | 170 | 55 |
| | Chicago | 8,475 | 204 | 86 |
| | London | 16,306 | 442 | 141 |
| | Los Angeles | 4,809 | 113 | 40 |
| | New York | 14,647 | 387 | 113 |
| | Paris | 9,986 | 398 | 84 |
| | Rome | 8,034 | 487 | 51 |
| | San Francisco | 11,326 | 252 | 79 |

### 4.7 Recommending Tourist Attractions

After training the model, given a query $q = (u, c)$, we first obtain the embedding of user $u$ and the embeddings of all the tourist attractions in city $c$, based on which (9) is exploited to obtain the visit probabilities, and finally recommend top $k$ tourist attractions in city $c$ to user $u$.

## 5 EXPERIMENTS

In this part, the experimental datasets, settings, and results are presented to evaluate the recommendation performance of MEAL. Specifically, MEAL is compared to its simplified variants and the state-of-the-art methods to prove its superiority. In addition, an example of tourist attraction recommendation is provided to illustrate the effectiveness of MEAL.

### 5.1 Datasets

The two datasets used in this paper were crawled from Flickr by using the public API[1]. Dataset1 consists of 699,896 geo-tagged photos that were taken in Beijing, Chengdu, Guangzhou, Hangzhou, Hong Kong, and Shanghai in China. Dataset2 [44] covers more than 7 million photos taken by 7,387 users at nine tour cities (i.e., Barcelona, Berlin, Chicago, London, Los Angeles, New York, Paris, Rome, and San Francisco) all over the world. Since the geotags of photos are very important for finding travel locations, we remove the photos without geotags. After that, there remain 84,400 photos and 1,432 users. Table 1 shows the statistics of the two datasets after tourist attraction extraction introduced in Section 4.1.

---

[1] https://www.flickr.com/services/api/

Table 2: The search space and the final choice of the hyper-parameters

| Hyper-parameters | Search space | Final choice on Dataset1 | Final choice on Dataset2 |
|---|---|---|---|
| $m_1$ | [0.1, 0.2, 0.3, 0.4, 0.5, 0.6, 0.7, 0.8, 0.9] | 0.1 | 0.1 |
| $m_2$ | [0.2, 0.3, 0.4, 0.5, 0.6, 0.7, 0.8, 0.9] | 0.2 | 0.2 |
| $m_3$ | [0.3, 0.4, 0.5, 0.6, 0.7, 0.8, 0.9] | 0.3 | 0.4 |
| $m_4$ | [0.1, 0.2, 0.3, 0.4, 0.5, 0.6, 0.7, 0.8, 0.9] | 0.1 | 0.2 |
| $m_5$ | [0.2, 0.3, 0.4, 0.5, 0.6, 0.7, 0.8, 0.9] | 0.2 | 0.4 |
| $m_6$ | [0.1, 0.2, 0.3, 0.4, 0.5, 0.6, 0.7, 0.8, 0.9] | 0.1 | 0.2 |
| $\omega$ | [5, 10, 15, 20, 25] | 10 | 15 |
| $\gamma$ | [5, 10, 15, 20, 25] | 15 | 20 |
| $\lambda_1$ | [0.00001, 0.00003, 0.0001, 0.0003, 0.001, 0.003, 0.01, 0.03, 0.1] | 0.001 | 0.01 |
| $\lambda_2$ | [0.00001, 0.00003, 0.0001, 0.0003, 0.001, 0.003, 0.01, 0.03, 0.1] | 0.0003 | 0.001 |

### 5.2 Experimental Settings

Following the settings of Majid et al. [8], we set the parameters for P-DBSCAN, and set visit duration threshold $t_{\text{thr}} = 6$ hours. We use the VGG16 pretrained on ImageNet to extract visual features.

Models are trained and evaluated on a server with one GPU (Nvidia GTX 1080 Ti). The code is released on GitHub[2]. Adam with learning rate 0.001 is used as the optimizer to train the model and the total training epoch is set as 200. We use NNI[3] to automatically select proper hyper-parameter settings. The search space and the final choice of the hyper-parameters are given in Table 2.

To evaluate the performance of MEAL, for each individual user who has visited at least three cities, we select two of his/her visited cities for validation and test, while the rest cities are used for training. Specifically, if a user has visited $r$ cities, $r(r-1)$ segments would be obtained.

Mean average precision (MAP) is employed to evaluate the recommendation performance, which is a widely used evaluation metric for recommender systems [43] and can be calculated by (29) and (30).

$$\text{MAP}@k = (\sum_{i=1}^{m} \text{AP}@k)/m \tag{29}$$

$$\text{AP}@k = (\sum_{i=1}^{k}(\sum_{j=1}^{i} f_j)/i)/k \tag{30}$$

where $m$ denotes the number of users who have visited at least three cities. $f_j$ equals 1 if the user has really visited the $j$-th tourist attraction in the recommendation list, otherwise $f_j$ equals 0.

Paired *t*-tests are used to determine whether the recommendation performance of MEAL and each compared method is significantly different when the significance level is 5%.

We have fully tuned the parameters of all the comparison methods to ensure fairness.

### 5.3 Variant Comparison

Seven simplified variants of MEAL: MEAL w/o visual similarity, MEAL-U, MEAL-L, MEAL-U/L, MEAL-U&L, MEAL-max, and MEAL-average are designed to explore the effectiveness of model components. MEAL w/o visual similarity does not take the visual similarity into account, and uses only $L_{\text{prediction}}$ to train the VGG16 model. MEAL-U, MEAL-L, MEAL-U/L, and MEAL-U&L replace $L_{\text{QUIN}}$ with corresponding triplet losses, which ensure that given a photo, its visual similarities with the photos taken by the same user are higher than those with the other photos, its visual similarities with the photos taken at the same tourist attraction are higher than those with the other photos, its visual similarities with the photos taken by the same user or taken at the same tourist attraction are higher than those with the other photos, and its visual similarities with the photos taken by the same user at the same tourist attraction are higher than those with the other photos, respectively. MEAL-max and MEAL-average eliminate the self-attention mechanism, and obtain the visual representations of users and tourist attractions via max pooling and average pooling the visual features of corresponding geo-tagged photos, respectively. In addition, we compare with variant MEAL w/o pretrain that trains the VGG16 from scratch to evaluate the effectiveness of the pretrained VGG16.

---

2 https://github.com/revaludo/MEAL
3 https://github.com/Microsoft/nni

Table 3: The performance of MEAL and its simplified variants (mean ± standard deviation), * indicates MEAL performs significantly better than the compared method

| Methods | MAP@5 on Dataset1 | MAP@5 on Dataset2 | MAP@10 on Dataset1 | MAP@10 on Dataset2 |
|---|---|---|---|---|
| MEAL w/o visual similarity | 0.217±0.067* | 0.228±0.039* | 0.195±0.036* | 0.207±0.043* |
| MEAL-U | 0.232±0.040* | 0.244±0.053* | 0.210±0.052* | 0.233±0.062* |
| MEAL-L | 0.245±0.063* | 0.261±0.081* | 0.224±0.057* | 0.246±0.028* |
| MEAL-U/L | 0.268±0.074* | 0.275±0.061* | 0.247±0.083* | 0.255±0.066* |
| MEAL-U&L | 0.283±0.035* | 0.298±0.072* | 0.261±0.028* | 0.273±0.038* |
| MEAL-max | 0.286±0.054* | 0.312±0.057* | 0.263±0.029* | 0.276±0.052* |
| MEAL-average | 0.292±0.042* | 0.321±0.046* | 0.270±0.086* | 0.281±0.054* |
| MEAL w/o pretrain | 0.303±0.038* | 0.336±0.060* | 0.281±0.049* | 0.291±0.036* |
| **MEAL** | **0.312±0.047** | **0.351±0.054** | **0.297±0.034** | **0.309±0.046** |

Table 3 shows the experimental results, from which the following observations can be concluded:

1) The recommendation performance of MEAL and its variants related to visual similarity are ranked as follows: MEAL w/o visual similarity < MEAL-U < MEAL-L < MEAL-U/L < MEAL-U&L < MEAL. MEAL w/o visual similarity performs the worst, as it does not consider any visual similarity level regarding users or tourist attractions that can reflect users' preferences or tourist attractions' characteristics. MEAL-L outperforms MEAL-U, which indicates that the variation of the photos taken at the same tourist attraction is smaller than that of the photos taken by the same user, as the scenery of a tourist attraction is rather stable, while the preference of a user may be diverse. MEAL-U/L outperforms MEAL-U and MEAL-L, as it considers both the visual similarity levels regarding users defined in MEAL-U and the visual similarity levels regarding tourist attractions defined in MEAL-L. MEAL-U&L outperforms MEAL-U/L, as it crosses the user and tourist attraction information to provide fine-grained visual similarity levels regarding both users and tourist attractions. MEAL outperforms MEAL-U&L, as it considers different ways of crossing the user and tourist attraction information to provide multi-level visual similarity.

2) MEAL outperforms MEAL-max and MEAL-average, which verifies the effectiveness of exploiting the self-attention mechanism to capture the significance of different photos for representing users and tourist attractions. In addition, MEAL-average slightly outperforms MEAL-max, which might be that MEAL-max is more vulnerable to outliers.

Table 4: The performance of MEAL and the compared methods (mean ± standard deviation), * indicates MEAL performs significantly better than the compared method

| Methods | MAP@5 on Dataset1 | MAP@5 on Dataset2 | MAP@10 on Dataset1 | MAP@10 on Dataset2 |
| --- | --- | --- | --- | --- |
| DTMMF | 0.228±0.059* | 0.232±0.046* | 0.207±0.074* | 0.213±0.061* |
| WIND-MF | 0.241±0.024* | 0.260±0.027* | 0.224±0.043* | 0.241±0.035* |
| VPMF | 0.256±0.061* | 0.276±0.047* | 0.231±0.068* | 0.250±0.056* |
| VPOI | 0.280±0.032* | 0.304±0.036* | 0.265±0.028* | 0.278±0.033* |
| **MEAL** | **0.312±0.047** | **0.351±0.054** | **0.297±0.034** | **0.309±0.046** |

3) MEAL outperforms w/o pretrain, which verifies the effectiveness of exploiting the pretrained VGG16 to extract visual features.

### 5.4 The Comparison of Different Methods

To show the superiority of MEAL, the state-of-the-art visual content based tourist attraction recommendation methods exploiting geo-tagged photos, i.e., DTMMF [10], WIND-MF [11], VPMF [12], and VPOI [13] are compared.

Table 4 shows the experimental results, from which the following observations can be concluded:

1) DTMMF performs the worst among all the methods. The reason might be that DTMMF only extracts the gender and age information in photos that contain people faces, ignoring other photos and a lot of other visual information.

2) VPMF performs better than WIND-MF. The reason might be that WIND-MF only extracts visual features via DNNs, while VPMF also extracts the traditional color histogram features and SIFT features.

3) VPOI performs better than VPMF. The reason might be that VPOI jointly extracts visual features and factorizes the user-tourist attraction interaction matrix for recommendation, while the extraction of visual features in VPMF has no relationship with the recommendation scenario.

4) MEAL performs better than VPOI. The reason might be that MEAL crosses the user and tourist attraction information to partition other photos into different groups for multi-level similarity aware visual content embedding. In addition, VPOI treats different photos taken by a user or taken at a tourist attraction equally, while MEAL exploits the self-attention mechanism to infer the weights of different photos for representing users and tourist attractions.

### 5.5 Case Study

To illustrate the effectiveness of our model, we give an example of recommending tourist attractions in Hangzhou to a user based on his/her travel history in Shanghai.

Table 5 shows the top 10 tourist attractions in Hangzhou recommended to the user and the user's true travel history in Shanghai and Hangzhou, from which the following observations can be concluded:

1) WIND-MF discovering four relevant tourist attractions, i.e., Zhejiang University, West Lake, Broken Bridge, and Lingyin Temple, ranking 2nd, 4th, 5th, and 8th in the recommendation results, respectively, performs better than DTMMF, which also discovers four relevant tourist attractions, i.e., West Lake, Leifeng Pagoda, Lingyin Temple, and Zhejiang University, ranking 3rd, 5th, 7th, and 8th in the recommendation results, respectively. The reason might be that WIND-MF exploits more visual information hidden in the geo-tagged photos compared to DTMMF, which only uses the gender and age information in photos that contain people faces.

Table 5: An example of tourist attraction recommendation results

| | |
|---|---|
| Visited tourist attractions in Shanghai | The Bund, the Huangpu River, Happy Valley, Town God's Temple, Yu Garden, People's Square, the Peace Hotel, Nanjing Road Pedestrian Street, Huaihai road, Oriental Pearl TV Tower, Garden Bridge of Shanghai, Tianzifang, Sinan Mansion, Xintiandi, Wukang Road. |
| Methods | Top 10 tourist attractions in Hangzhou recommended to the user |
| DTMMF | The Grand Canal, China National Silk Museum, **West Lake**, Yuefei Temple, **Leifeng Pagoda**, Hefang Street, **Lingyin Temple**, **Zhejiang University**, Hupao Spring, Hangzhou Zoo. |
| WIND-MF | Hang Zhou Botanical Garden, **Zhejiang University**, Hangzhou Museum, **West Lake**, **Broken Bridge**, Children's Palace, Hefang Street, **Lingyin Temple**, Hupao Spring, Hangzhou Zoo. |
| VPMF | **Broken Bridge**, Bai Causeway, **West Lake**, **HangZhou Polar Ocean Park**, China National Silk Museum, **Zhejiang University**, Hang Zhou Botanical Garden, **Lingyin Temple**, Hangzhou Taiziwan Park, Xixi National Wetland Park. |
| VPOI | Hangzhou Museum, **West Lake**, Su Causeway, **Kushan**, **Zhejiang University**, Xixi National Wetland Park, **the Qiantang River**, Longjing Village, **Meijiawu**, **Faxi Temple**. |
| MEAL | **Zhejiang University**, **West Lake**, Hangzhou Museum, Three Pools Mirroring the Moon, **the Qiantang River**, **Hangzhou Paradise**, Baochu pagoda, **Leifeng Pagoda**, Hangzhou Taiziwan Park, **Lingyin Temple**. |
| Visited tourist attractions in Hangzhou | Zhejiang University, West Lake, Broken Bridge, Kushan, Hangzhou Paradise, Xiang Lake, Leifeng Pagoda, Lingyin Temple, Southern Song Imperial Street, the Qiantang River, Pagoda of Six Harmonies, Meijiawu, Faxi Temple, HangZhou Polar Ocean Park. |

2) VPMF discovers more relevant tourist attractions than WIND-MF. The reason might be that VPMF extracts more visual features from the geo-tagged photos to represent users and tourist attractions.

3) VPOI discovers more relevant tourist attractions than VPMF, which indicates that jointly modeling the visual content embedding and recommendation tasks can extract visual features adaptive to tourist attraction recommendation.

4) MEAL discovering six relevant tourist attractions, i.e., Zhejiang University, West Lake, the Qiantang River, Hangzhou Paradise, Leifeng Pagoda, and Lingyin Temple, ranking 1st, 2nd, 5th, 6th, 8th and 10th in the recommendation results, respectively, performs better than VPOI, which also discovers six relevant tourist attractions, i.e., West Lake, Kushan, Zhejiang University, the Qiantang River, Meijiawu, and Faxi Temple, ranking 2nd, 4th, 5th, 7th, 9th, and 10th in the recommendation results, respectively. These results demonstrate the effectiveness of multi-level similarity aware visual content embedding and self-attention mechanism based visual representation learning.

### 5.6 Convergence

In order to evaluate the impact of the hard mining technique on the convergence of training, we give the training loss of the proposed method in Figure 5. As shown in the figure, the training loss first decreases gradually and tends to be stable, which illustrates the convergence property of the propose method.

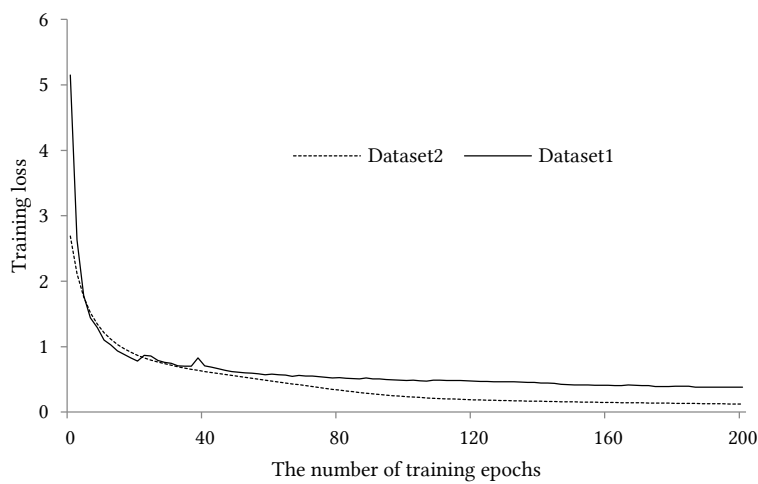

Figure 5: The training loss of the proposed method.

## 6 CONCLUSIONS AND FUTURE WORK

In this paper, multi-level visual similarity based personalized tourist attraction recommendation using geo-tagged photos (MEAL) is proposed. MEAL embeds the visual contents of photos by exploiting a quintuplet loss to ensure the proper order of the visual similarities defined by crossing the user and tourist attraction information, represents users and tourist attractions by exploiting the self-attention mechanism to infer the weights of different photos as well as by factorizing the user-tourist attraction interaction matrix, and predicts the visit probabilities for recommendation. We conducted experiments on two datasets crawled from Flickr, and the experimental results proved the advantage of this method.

The main limitation of the proposed method is that the sequence information that can be captured from the taken time of photos is ignored when we obtain the latent factors and the visual representations of users and travel locations. For further expansion of the proposed method, we consider exploiting the photo sequences of users and tourist attractions to better represent them. In addition, we will introduce the categories of tourist attractions to provide more visual similarity levels.